\documentstyle[12pt,epsfig]{article}
\expandafter\ifx\csname mathrm\endcsname\relax\def\mathrm#1{{\rm #1}}\fi

\makeatletter

\newcount\@tempcntc
\def\@citex[#1]#2{\if@filesw\immediate\write\@auxout{\string\citation{#2}}\fi
  \@tempcnta\z@\@tempcntb\m@ne\def\@citea{}\@cite{\@for\@citeb:=#2\do
    {\@ifundefined
       {b@\@citeb}{\@citeo\@tempcntb\m@ne\@citea
        \def\@citea{,\penalty\@m\ }{\bf ?}\@warning
       {Citation `\@citeb' on page \thepage \space undefined}}%
    {\setbox\z@\hbox{\global\@tempcntc0\csname
b@\@citeb\endcsname\relax}%
     \ifnum\@tempcntc=\z@ \@citeo\@tempcntb\m@ne
       \@citea\def\@citea{,\penalty\@m}
       \hbox{\csname b@\@citeb\endcsname}%
     \else
      \advance\@tempcntb\@ne
      \ifnum\@tempcntb=\@tempcntc
      \else\advance\@tempcntb\m@ne\@citeo
      \@tempcnta\@tempcntc\@tempcntb\@tempcntc\fi\fi}}\@citeo}{#1}}

\def\@citeo{\ifnum\@tempcnta>\@tempcntb\else\@citea
  \def\@citea{,\penalty\@m}%
  \ifnum\@tempcnta=\@tempcntb\the\@tempcnta\else
   {\advance\@tempcnta\@ne\ifnum\@tempcnta=\@tempcntb \else
\def\@citea{--}\fi
    \advance\@tempcnta\m@ne\the\@tempcnta\@citea\the\@tempcntb}\fi\fi}

\makeatother

\def\asymp#1%
{\mathrel{\raisebox{-.4em}{$\widetilde{\scriptstyle #1}$}}}
\def\Nlim#1{\mathrel{\raisebox{-.4em}
{$\stackrel{\disp\longrightarrow}{\scriptstyle#1}$}}}
\def\Nequal#1%
{\mathrel{\raisebox{-.5em}{$\stackrel{=}{\scriptstyle\rm#1}$}}}

\def\beq{\begin{equation}}
\def\eeq{\end{equation}}
\def\beqar{\begin{eqnarray}}
\def\eeqar{\end{eqnarray}}
\def\barr#1{\begin{array}{#1}}
\def\earr{\end{array}}
\def\bfi{\begin{figure*}}
\def\efi{\end{figure*}}
\def\btab{\begin{table}}
\def\etab{\end{table}}
\def\bce{\begin{center}}
\def\ece{\end{center}}

\def\disp{\displaystyle}
\def\text{\textstyle}
\def\fs{\footnotesize}

\def\al{\alpha}
\def\be{\beta}

\def\ga{\gamma}
\def\de{\delta}
\def\De{\Delta}

\def\si{\sigma}

\def\refeq#1{\mbox{(\ref{#1})}}

\def\reffi#1{\mbox{Fig.~\ref{#1}}}

\def\refta#1{\mbox{Table~\ref{#1}}}

\def\citere#1{\mbox{Ref.~\cite{#1}}}
\def\citeres#1{\mbox{Refs.~\cite{#1}}}
 
\newcommand{\TeV}{\unskip\,\mathrm{TeV}}
\newcommand{\GeV}{\unskip\,\mathrm{GeV}}

\newcommand{\pb}{\unskip\,\mathrm{pb}}
\newcommand{\fb}{\unskip\,\mathrm{fb}}

\newcommand{\rd}{{\mathrm{d}}}
\newcommand{\ri}{{\mathrm{i}}}
\newcommand{\rU}{{\mathrm{U}}}
\newcommand{\rL}{{\mathrm{L}}}
\newcommand{\rT}{{\mathrm{T}}}

\newcommand{\Oa}{\mathswitch{{\cal{O}}(\alpha)}}

\def\mathswitchr#1{\relax\ifmmode{\mathrm{#1}}\else$\mathrm{#1}$\fi}

\newcommand{\PW}{\mathswitchr W}
\newcommand{\PZ}{\mathswitchr Z}

\newcommand{\PH}{\mathswitchr H}

\newcommand{\Pt}{\mathswitchr t}

\newcommand{\Pep}{\mathswitchr {e^+}}
\newcommand{\Pem}{\mathswitchr {e^-}}
\newcommand{\PWp}{\mathswitchr {W^+}}
\newcommand{\PWm}{\mathswitchr {W^-}}

\def\mathswitch#1{\relax\ifmmode#1\else$#1$\fi}

\newcommand{\MW}{\mathswitch {M_\PW}}

\newcommand{\MZ}{\mathswitch {M_\PZ}}
\newcommand{\MH}{\mathswitch {M_\PH}}

\newcommand{\Mt}{\mathswitch {m_\Pt}}

\def\ie{i.e.\ }
\def\eg{e.g.\ }

\newcommand{\ct}{counterterm}

\newcommand{\cs}{cross-section}
\newcommand{\css}{cross-sections}

\newcommand{\mpar}[1]{{\marginpar{\hbadness10000%
                      \sloppy\hfuzz10pt\boldmath\bf#1}}%
                      \typeout{marginpar: #1}\ignorespaces}
\newcommand{\AAWW}{\gamma\gamma\to\PWp\PWm}

\newcommand{\eeWW}{\Pep\Pem\to\PWp\PWm}
\newcommand{\eAnW}{\Pem\gamma\to\PWm\nu}
\newcommand{\Born}{\mathrm{Born}}

\newcommand{\bos}{\mathrm{bos}}
\newcommand{\ferm}{\mathrm{ferm}}
\newcommand{\Coul}{\mathrm{Coul.}}

\newcommand{\cut}{\mathrm{cut}}

\newcommand{\unpol}{\mathrm{unpol}}

\def\draftdate{\relax}
\def\mda{\relax}
\def\mua{\relax}
\def\mla{\relax}
\def\draft{
\def\thtystars{******************************}
\def\sixtystars{\thtystars\thtystars}
\typeout{}
\typeout{\sixtystars**}
\typeout{* Draft mode!
         For final version remove \protect\draft\space in source file *}
\typeout{\sixtystars**}
\typeout{}
\def\draftdate{\today}
\def\mda{\mpar{$\downarrow$}}
\def\mua{\mpar{$\uparrow$}}
\def\mla{\marginpar[\boldmath\hfil$\rightarrow$\hfil]%
                   {\boldmath\hfil$\leftarrow $\hfil}%
                    \typeout{marginpar:
$\leftrightarrow$}\ignorespaces}
\def\mua{\marginpar[\boldmath\hfil$\uparrow$]%
                   {\boldmath$\uparrow$\hfil}%
                    \typeout{marginpar: $\uparrow$}\ignorespaces}
\def\mda{\marginpar[\boldmath\hfil$\downarrow$]%
                   {\boldmath$\downarrow$\hfil}%
                    \typeout{marginpar: $\downarrow$}\ignorespaces}
\def\mla{\marginpar[\boldmath\hfil$\rightarrow$]%
                   {\boldmath$\leftarrow $\hfil}%
                    \typeout{marginpar: $\leftrightarrow$}\ignorespaces}
\overfullrule 5pt
\oddsidemargin -15mm
\marginparwidth 29mm
}

\def\rem#1{{\em #1}}
\def\rem#1{{#1}}

\begin{document}

\thispagestyle{empty}
\def\thefootnote{\fnsymbol{footnote}}
\setcounter{footnote}{1}
\null
\strut\hfill  BI-TP 96/03 \\
\strut\hfill WUE-ITP-96-001\\
\strut\hfill hep-ph/9601355
\vskip 0cm
\vfill
\begin{center}
{\Large \bf 
 \boldmath{Electroweak Radiative Corrections to $\gamma\gamma\to\PWp\PWm$}
\unboldmath%
\footnote{Contribution to the Proceedings of the Workshop on
{\it Physics with $e^+e^-$-Colliders,}
Annecy, Gen\`eve, Hamburg, February 4 to September 1, 1995.}
\par} \vskip 2.5em
{\large
{\sc A.~Denner }\\[1ex]
{\normalsize \it Institut f\"ur Theoretische Physik, Universit\"at W\"urzburg\\
Am Hubland, D-97074 W\"urzburg, Germany}
\\[2ex]
{\sc S.~Dittmaier }\\[1ex]
{\normalsize \it Theoretische Physik, Universit\"at Bielefeld\\ 
Postfach 100131, D-33501 Bielefeld, Germany}
\\[2ex]
{\sc R.~Schuster%
\footnote{Supported by the Deutsche Forschungsgemeinschaft.}
} \\[1ex]
{\normalsize \it Institut f\"ur Theoretische Physik, Universit\"at W\"urzburg\\
Am Hubland, D-97074 W\"urzburg, Germany}
}
\par \vskip 1em
\end{center} \par
\vskip 1cm 
\vfill
{\bf Abstract:} \par
We discuss the complete virtual and soft-photonic ${\cal O}(\alpha)$
corrections to $\gamma\gamma\to\PW^+\PW^-$ within the electroweak
Standard Model for arbitrary polarized photons.
In the on-shell renormalization scheme for fixed $\MW$
no leading corrections associated with the running of $\alpha$ or
heavy top-quark and Higgs-boson masses occur.
The corrections turn out to be of the order of 10\%, but can become much
larger where the lowest-order cross-sections are suppressed.
\par
\vskip 1cm 
\noindent BI-TP 96/03 \\
WUE-ITP-96-001\par
\vskip .15mm
\noindent January 1996 \par
\null
\setcounter{page}{0}
\clearpage

\def\thefootnote{\arabic{footnote}}
\setcounter{footnote}{0}

\def\thefootnote{\fnsymbol{footnote}}
\setcounter{footnote}{1}
\begin{center}
{\Large \bf 
 Electroweak Radiative Corrections to \boldmath{$\gamma\gamma\to\PWp\PWm$} }

\vspace*{0.3cm}

A.\ Denner$^1$, S.~Dittmaier$^2$ and R. Schuster$^{1,}$\footnotemark \\[2.5ex]
 \parbox{12.8cm}{
 {\it$^1$ Institut f\"ur Theoretische Physik, 
          Universit\"at W\"urzburg, Germany}  \\[1ex]
 {\it$^2$ Theoretische Physik, Universit\"at Bielefeld, Germany} } 
\end{center} 
 \baselineskip=18pt

\vspace*{1cm}

\footnotetext{Supported by the Deutsche Forschungsgemeinschaft.}
\renewcommand{\thefootnote}{\arabic{footnote}}
\setcounter{footnote}{0}

\section{Introduction}

A particularly interesting process at $\ga\ga$ colliders is 
\PW-pair production.
Its total \cs\ approaches a constant of about
$80\pb$ at high energies 
corresponding to $8\times10^5$ \PW~pairs for $10\fb^{-1}$.
Although this large \cs\ 
is drastically reduced by angular cuts, even for $|\!\cos\theta|< 0.8$
it is still 15 and $4\pb$ at a center-of-mass energy of 500 and
$1000\GeV$, respectively, and thus much larger as \eg the one for $\eeWW$.
Hence $\AAWW$ is very well-suited for precision investigations of the SM.

Most of the existing works on $\AAWW$ concentrated on tree-level 
predictions and
on the influence of anomalous non-Abelian gauge couplings \cite{Ki73,Ye91,Be92}.
At tree level, the process $\AAWW$  depends both on the triple
$\ga\PW\PW$ and the quartic $\ga\ga\PW\PW$ coupling, and
no other vertices are involved in the unitary gauge.
The sensitivity to the $\ga\PW\PW$ coupling is comparable and 
complementary to the reactions $\eeWW$ and $\eAnW$
\cite{Ye91}.
Because the sensitivity to the $\ga\ga\PW\PW$ coupling is much larger 
than the one in $\Pep\Pem$ processes, $\AAWW$ is the ideal process to study 
this coupling \cite{Be92}. 

The one-loop diagrams involving a resonant Higgs boson have been
calculated in order to 
study the possible investigation of the Higgs boson via
$\ga\ga\to\PH^*\to\PWp\PWm$ \cite{Va79,Bo92,Mo94,Ve94}.
Based on our complete one-loop calculation \cite{aaww}, we have
supplemented these investigations by a discussion of the heavy-Higgs
effects in \citere{HH}. As a matter of fact, only the (suppressed)
channels of longitudinal \PW-boson production are sensitive to the Higgs
mechanism, but the (dominant) channels of purely transverse \PW-boson 
production are extremely insensitive. 
This insensitivity to the Higgs sector renders $\AAWW$ even more
suitable for the investigation of the non-Abelian self couplings.

In this paper we summarize our results for the {\em complete} virtual and
soft-photonic $\Oa$ corrections. We give a survey of the leading
corrections and restrict the numerical discussion to unpolarized
$\PW$ bosons.  More detailed results and a discussion of their evaluation
can be found in \citeres{aaww,HH}.

\section{Lowest-order \cs}
\label{se:born}

The Born \cs\ of $\AAWW$ is given by the formulae%
\beq
\left(\frac{\rd\sigma}{\rd\Omega}\right)^{\Born}_{\unpol} =
 \frac{3 \alpha^2 \beta}{2s} 
\Biggl\{1 - \frac{2s(2s+3\MW^2)}{3(\MW^2 -t)(\MW^2 -u)} 
+ \frac{2s^2(s^2+3\MW^4)}{3(\MW^2 -t)^2(\MW^2 -u)^2} 
   \Biggr\} 
\label{unpol}
\eeq
for the unpolarized differential \cs\ and 
\beq
\sigma^{\Born}_{{\unpol}} = \frac{6\pi \alpha^2 }{s}
\beta \cos\theta_{{\cut}}
\Biggl\{
1 - 
\frac{4\MW^2(s-2\MW^2)}{s^2\beta \cos\theta_{{\cut}} }
\log\left(\frac{1 + \beta \cos\theta_{{\cut}}}
{1 - \beta \cos\theta_{{\cut}}}\right)
+ \frac{16(s^2+3\MW^4)}{3s^2
(1 - \beta^2\cos^2\theta_{\cut})} \Biggr\} 
\eeq
for the unpolarized \cs\ integrated over
$\theta_{\cut}<\theta<180^\circ-\theta_{\cut}$.
Here $\beta=\sqrt{1-4\MW^2/s}$ denotes the velocity of the W bosons and
$s$, $t$, and $u$ are the usual Mandelstam variables.
Concerning kinematics, polarizations, input parameters, etc., we
follow 
the conventions of \citere{aaww} throughout.

As can be seen from \refta{table_born}, the lowest-order \css\ are
dominated by transverse (T) W~bosons. The massive $t$-channel exchange
gives rise to a constant \cs\ at high energies, $s \gg \MW^2$, 
for $\theta_{{\cut}} = 0$
\beq
\sigma^{\Born}_{\pm\pm\rU\rU}, \sigma^{\Born}_{\pm\mp\rU\rU}
\;\;\Nlim{s\rightarrow\infty}\;\;
\sigma^{\Born}_{\pm\pm\rT\rT}, \sigma^{\Born}_{\pm\mp\rT\rT}
\;\;\Nlim{s\rightarrow\infty}\;\;
\frac{8 \pi \alpha^2}{\MW^2} = 80.8 \pb.
\eeq
For a finite cut, $\sigma^{\Born}_{\pm\pm\rU\rU}$  and
$\sigma^{\Born}_{\pm\mp\rU\rU}$ behave as $1/s$ for large $s$.
This is illustrated in \reffi{fi:intcs10} for
two different angular cuts $\theta_{\cut} = 10^\circ, 20^\circ$.
Close to threshold the differential and integrated 
\css\ for all polarization configurations vanish like $\beta$.
\begin{table}
\footnotesize
\begin{center}
\arraycolsep 6pt
$$\begin{array}{|c|c||c|c|c|c|c|c|}
\hline
\sqrt{s}/\mathrm{GeV}
 & \theta &
{\mathrm{unpol}} & {{\pm\pm}\mathrm{TT}} & {{\pm\pm}\mathrm{LL}} &
{{\pm\mp}\mathrm{TT}} & {{\pm\mp}\mathrm{LL}} & {{\pm\mp}\mathrm{(LT+TL)}} \\
\hline\hline
\phantom{0}500 & \phantom{0}0^\circ<\theta<180^\circ & 
77.6 &  82.2 & 6.10\times 10^{-2} & 70.2 & 9.99 \times 10^{-1} & 
1.69 \phantom{{}\times 10^{-1}} \\
\cline{2-8}
& 20^\circ<\theta<160^\circ & 
36.7 & 42.7 & 3.17\times 10^{-2}& 28.2 & 9.89\times 10^{-1}& 
1.49\phantom{{}\times 10^{-1}} \\
\hline\hline
1000 & \phantom{0}0^\circ<\theta<180^\circ & 
80.1 & 82.8 & 3.54 \times 10^{-3} & 76.9 & 2.52\times 10^{-1} & 
1.70\times 10^{-1}\\
\cline{2-8}
& 20^\circ<\theta<160^\circ & 
14.2 & 16.8 &7.18 \times 10^{-4}& 11.2 & 2.44\times 10^{-1} & 
1.21\times 10^{-1}\\
\hline\hline
2000 & \phantom{0}0^\circ<\theta<180^\circ & 
80.6 & 81.6 & 2.14\times 10^{-4}& 79.5 & 6.41\times 10^{-2}& 
1.50\times 10^{-2}\\
\cline{2-8}
& 20^\circ<\theta<160^\circ & 
4.07& 4.84 & 1.27\times 10^{-5}& 3.23 & 6.11\times 10^{-2}& 
8.26\times 10^{-3}  \\
\hline
\end{array}$$
\caption{\em Lowest-order integrated \css\ in pb for several polarizations;
the lowest-order \cs\ for ${\pm\pm}\mathrm{(LT+TL)}$ vanishes.}
\label{table_born}
\end{center}
\end{table}

\section{Leading corrections}
\label{leadrcs}

Electroweak radiative corrections contain leading contributions of 
universal origin.
In the on-shell renormalization scheme with input parameters $\al$, $\MW$, 
$\MZ$ these affect the corrections to $\AAWW$ as follows:
\begin{itemize}
\item
Since the {\it two} external photons are on mass shell,
the relevant effective coupling is the one at zero-momentum transfer
and the running of $\al$ is not relevant.
\item
In order to handle the Higgs-boson resonance at $s=\MH^2$,
in the literature \cite{Bo92,Mo94,Ve94} the Higgs-boson width has been
introduced na{\"\i}vely by the replacement
\beq
\frac{F^H(s)}{s-\MH^2} \;\;\longrightarrow\;\; 
\frac{F^H(s)}{s-\MH^2 + \ri \MH\Gamma_{\PH}} 
\eeq
in the resonant contribution.
However, this treatment destroys gauge invariance at the level of 
non-resonant $\Oa$ corrections.
In order to preserve gauge invariance, we decompose the 
Higgs-resonance contribution into a gauge-invariant resonant part and a
gauge-dependent non-resonant part and introduce $\Gamma_\PH$ only
in the former:
\beq
\frac{F^H(s)}{s-\MH^2} \;\;\longrightarrow\;\;
\frac{F^H(\MH^2)}{s-\MH^2 + \ri \MH\Gamma_{\PH}}
+ \frac{F^H(s) - F^H(\MH^2)}{s-\MH^2} \;.
\label{eq_higgs2}
\eeq
\item
Outside the region of 
Higgs resonance the Higgs-mass dependence is small for unpolarized \PW~bosons.
\looseness -1
For all polarizations no corrections involving
$\log(\MH/\MW)$ or $\MH^2/\MW^2$ arise in the heavy-Higgs limit
\cite{HH}.
However, for $\sqrt{s}\gg\MH\gg\MW$ corrections
proportional to $\MH^2/\MW^2$ appear for the \css\ involving
longitudinal ($\rL$)
gauge bosons as a remnant of the unitarity cancellations.
These terms give rise to large effects in
particular for \css\ with longitudinal \PW~bosons.
\item
The top-mass-dependent 
corrections are also small and behave
similar to the Higgs-mass-dependent corrections for $\sqrt{s}\gg\Mt\gg\MW$;
more precisely neither corrections proportional to $\Mt^2$ nor proportional
to $\log\Mt$ occur
in this limit for fixed $\MW$.
\item
As $\AAWW$ involves no light charged external particles,
no large logarithmic corrections associated with collinear
photons show up (apart from the region of very high energies, $s\gg\MW^2$).
As a consequence, the photonic corrections are not enhanced with respect
to the weak corrections.
\item
Close to threshold, \ie for $\be\ll 1$, the Coulomb singularity gives rise to 
the large universal correction
\beq
\de\si^{\Coul} = \frac{\al\pi}{2\be} \si^{\Born}.
\label{eq:coul}
\eeq
The factor $\beta^{-1}$ is 
typical for the pair production of stable (on-shell) particles.
For the generalization to unstable (off-shell) particles see \citere{coul}.
\end{itemize}

At high energies, $s\gg\MW^2$, the radiative corrections are dominated 
by terms like
$(\al/\pi)\log^2(s/\MW^2)$, which arise from vertex and box diagrams.
At $1\TeV$ these are about 10\%, setting the scale for the (weak)
radiative corrections at this energy.

\section{Numerical results}
\label{se:num}

Electromagnetic and genuine weak corrections cannot be separated in a
gauge-invariant way 
on the basis of Feynman diagrams.
As no leading collinear logarithms occur in 
$\AAWW$, the only source of enhanced photonic corrections are
the soft-photon-cut-off-dependent terms which yield the relative correction
\beq
\de_{\cut} = -\frac{2\al}{\pi} \log\left(\frac{\De E}{E}\right) 
\left(1 - \frac{s-2\MW^2}{s\be}\log\frac{1+\be}{1-\be}\right).
\label{eq:cut}
\eeq
Since we are mainly interested in the 
weak corrections, 
we discard the (gauge-invariant)
cut-off-dependent terms \refeq{eq:cut} by setting the soft-photon
cut-off energy $\De E$ equal to the beam energy $E$
and consider the rest as a suitable measure of
the weak corrections.
\rem{If not stated otherwise, the
correction $\de$ stands in the following for the complete 
relative soft-photonic and virtual electroweak corrections with $\De E = E$.}

\begin{figure*}
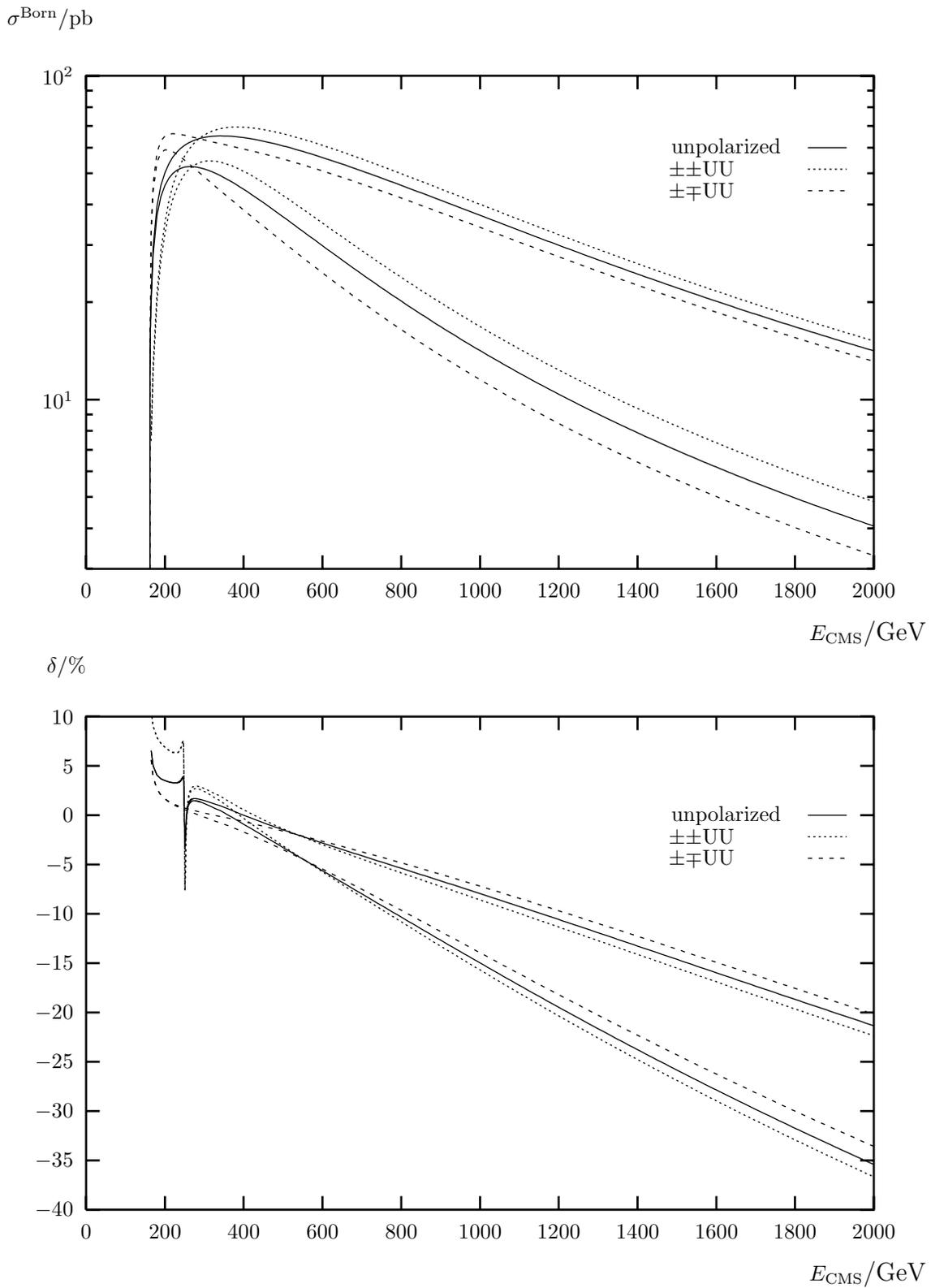

\setlength{\unitlength}{1mm}
\begin{picture}(160,210)(0,0)
\put(0,110){\input{aaww0.tot2.tex}}
\put(0,05){\input{aaww.tot2.tex}}
\end{picture}
\caption{\em Integrated lowest-order \css\ and corresponding relative 
corrections for an angular cut 
         $10^\circ \leq \theta \leq 170^\circ$ 
(upper set of curves in each plot) and
         $20^\circ \leq \theta \leq 160^\circ$
(lower set).}
\label{fi:intcs10}
\end{figure*}

Figure \ref{fi:intcs10} shows the 
``weak'' corrections to the total \css\ 
integrated over $10^\circ \leq \theta \leq 170^\circ$ and
$20^\circ \leq \theta \leq 160^\circ$ for unpolarized W~bosons.
The corrections for different photon polarizations almost coincide with 
each other
and reach roughly $-20\%$ for $\theta_{\cut} = 10^\circ$ and 
$-35\%$ for $\theta_{\cut} = 20^\circ$ at $\sqrt{s}=2\TeV$.
At low energies the \css\ with equal photon and \PW~boson helicities
are dominated by the Higgs resonance.
Note that owing to helicity conservation the other \css\ 
are not affected by the Higgs resonance.

\begin{table}
\footnotesize
\newdimen\digitwidth
\setbox0=\hbox{0}
\digitwidth=\wd0
\catcode`!=\active 
\def!{\kern\digitwidth}
\newdimen\minuswidth
\setbox0=\hbox{$-$}
\minuswidth=\wd0
\catcode`?=\active 
\def?{\kern\minuswidth}
\begin{center}
\arraycolsep 6pt
$$\begin{array}{|c|c||c|c|c|c|c|c|}
\hline
\sqrt{s}/\mathrm{GeV} & \theta & \sigma^{\mathrm{Born}}/\mathrm{pb} &
\delta_{\Delta E = 0.1E} /\% & 
\delta_{\cut} /\% & 
\delta_{\Delta E = E} /\% & 
\delta_{\bos}/\% &
\delta_{\ferm}/\% \\
\hline\hline
       & !5^\circ &  98.13  & !?0.02 &  -2.79 & ?!2.81 & ?!1.49 &  ?1.32 \\
 \cline{2-8}
       & 20^\circ &  26.04  & !-2.68 &  -2.79 & ?!0.11 & !-0.08 &  ?0.19 \\
\cline{2-8}
  !500 & 90^\circ &  0.724  & -10.79 &  -2.79 & !-8.00 & !-5.62 &  -2.38 \\
\cline{2-8}
       & !0^\circ< \theta <180^\circ &  77.55     & !-3.38 &   -2.79 
& !-0.59 & !-0.65 &   ?0.06  \\
\cline{2-8}
       & 10^\circ< \theta <170^\circ &  60.74     & !-4.27 &   -2.79 &
  !-1.48 & !-1.21 &   -0.27  \\
\cline{2-8}
       & 20^\circ< \theta <160^\circ &  36.67     & !-6.06 &  -2.79 
& !-3.27 & !-2.39 &   -0.89 \\
\hline\hline
       & !5^\circ &  291.9  & !-2.06 &  -4.31 & ?!2.25 & ?!1.04 &  ?1.21 \\
\cline{2-8}
       & 20^\circ &  15.61  & -11.90 &  -4.31 & !-7.59 & !-6.37 &  -1.22 \\
\cline{2-8}
 1000  & 90^\circ &  0.193  & -31.64 &  -4.31 & -27.33 & -21.93 &  -5.40 \\
\cline{2-8}
       & !0^\circ< \theta <180^\circ &  80.05    & !-7.08 &  -4.31 
& !-2.77 & !-2.71 &   -0.06 \\ 
\cline{2-8}
       & 10^\circ< \theta <170^\circ &  37.06    & -12.26 &  -4.31 
& !-7.95 & !-6.65 &   -1.30 \\
\cline{2-8}
       & 20^\circ< \theta <160^\circ & 14.16     & -19.29 &  -4.31 
& -14.98 & -12.20 &  -2.78 \\
\hline\hline
       & !5^\circ &  418.8 & !-7.14 &  -5.80 & !-1.33 & !-1.59 &  ?0.25 \\
\cline{2-8}
       & 20^\circ &  5.163 & -30.31 &  -5.80 & -24.51 & -20.96 &  -3.55 \\
\cline{2-8}
 2000  & 90^\circ &  0.049 & -59.59 &  -5.80 & -53.78 & -45.47 &  -8.32 \\ 
\cline{2-8}
       & !0^\circ< \theta <180^\circ &  80.59     & !-9.85 &  -5.80 
& !-4.04 & !-3.95 &  -0.09 \\
\cline{2-8}
       & 10^\circ< \theta <170^\circ &  14.14     & -27.15 &  -5.80 
& -21.35 & -18.34 &  -3.01 \\
\cline{2-8}
       & 20^\circ< \theta <160^\circ &  4.068     & -41.22 &  -5.80 
& -35.41 & -30.12 &  -5.29 \\
\hline
\end{array}$$
\caption{\em Lowest-order \css\ and relative corrections for unpolarized
particles}
\label{ta:born}
\end{center}
\end{table}
In \refta{ta:born} we list the unpolarized 
\css\ 
and the corresponding 
corrections for several energies and scattering angles. We include the
corrections for a soft-photon-energy cut-off $\Delta E = 0.1E$, i.e.\
the 
cut-off-dependent corrections $\de_{\cut}$ from \refeq{eq:cut}, and the 
individual (gauge-invariant) fermionic $\delta_{\ferm}$ 
and bosonic corrections  $\delta_{\bos}$.
The fermionic corrections consist of all loop diagrams and \ct\ 
contributions involving fermion loops, all other contributions form the 
bosonic corrections.
The fermionic corrections stay below 5--10\% even for high energies. 
The bosonic contributions are responsible for the
large corrections at high energies, in
particular in the central angular region. 
 
In \citere{Ye91} the total \cs\ and the ratios
\beqar
R_{\mathrm{IO}}=\frac{\si(|\!\cos\theta|<0.4)}{\si(|\!\cos\theta|<0.8)}, 
\qquad
R_{\mathrm{LT}}=\frac{\si_{\rL\rL}}{\si_{\rT\rT}}, 
\qquad
R_{\mathrm{02}}=\frac{\si_{++}}{\si_{+-}} .
\eeqar
have been investigated in view of their sensitivity to anomalous couplings.%
\footnote{Note that we do not perform a convolution with a realistic
photon spectrum but consider the incoming photons as monochromatic.}
We list the lowest-order predictions together with the $\Oa$-corrected
ones and the relative 
``weak'' corrections for these observables in
\refta{ta:obs} using $|\!\cos\theta_{\cut}|=0.8$.
\begin{table}
\newdimen\digitwidth
\setbox0=\hbox{0}
\digitwidth=\wd0
\catcode`!=\active 
\def!{\kern\digitwidth}
\newdimen\minuswidth
\setbox0=\hbox{$-$}
\minuswidth=\wd0
\catcode`?=\active 
\def?{\kern\minuswidth}
\begin{center}
\arraycolsep 6pt
$$\begin{array}{|c|c||c|c|c|c|}
\hline
\sqrt{s}/\mathrm{GeV} & & \sigma/\mathrm{pb} &
R_{\mathrm{IO}} & R_{\mathrm{LT}} & R_{02} \\
\hline\hline
       & \mathrm{Born~level}  & ?15.74  & ?0.265  &  0.0308  & ?1.934 \\
\cline{2-6}
 !500  & \mathrm{corrected}   & ?14.82  & ?0.259  &  0.0325  & ?1.950 \\
\cline{2-6}
       & \mathrm{corrections/\%} & !-5.83  & !-2.02  &  !5.43!  & ?0.78! \\
\hline\hline
       & \mathrm{Born~level}  & ?4.659  & ?0.241  &  0.0235  & ?2.229 \\
\cline{2-6}
 1000  & \mathrm{corrected}   & ?3.617  & ?0.227  &  0.0276  & ?2.184 \\
\cline{2-6}
       & \mathrm{corrections/\%} & -22.36  & !-5.64  &  17.08!  & -2.05! \\
\hline\hline
       & \mathrm{Born~level}  & ?1.218  & ?0.234  &  0.0220  & ?2.307 \\
\cline{2-6}
 2000  & \mathrm{corrected}   & ?0.647  & ?0.207  &  0.0321  & ?2.168 \\
\cline{2-6}
       & \mathrm{corrections/\%} & -46.86  &  -11.53  &  46.11!  & -6.02!\\
\hline
\end{array}$$
\caption{\em Tree-level and $\Oa$ results
for various observables using $|\!\cos\theta_{\cut}|=0.8$}
\label{ta:obs}
\end{center}
\end{table}

\begin{figure*}
\setlength{\unitlength}{1mm}
\begin{picture}(160,110)(0,0)
\put(0,10){\input{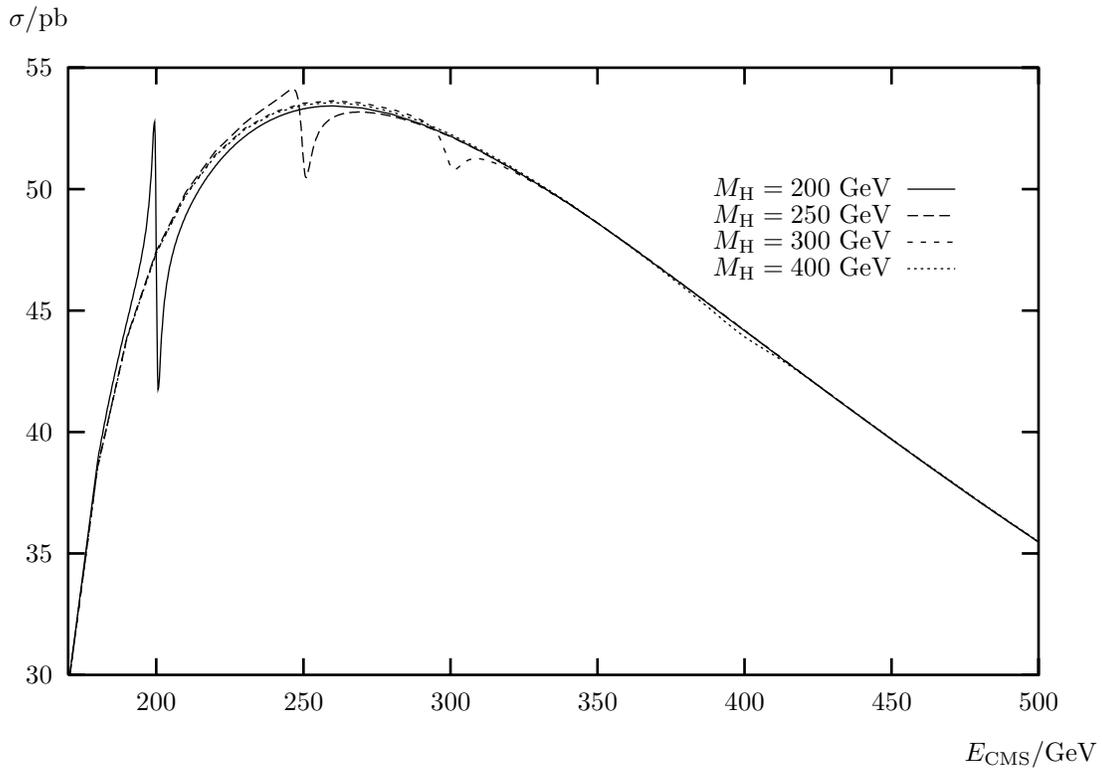}}
\end{picture}
\caption{\em Integrated unpolarized \cs\ including $\Oa$ ``weak'' corrections
for various Higgs-boson masses ($20^\circ<\theta<160^\circ$)}
\label{fi:intcshiggs}
\end{figure*}

In \reffi{fi:intcshiggs} we plot the integrated \cs\ including \Oa\ 
\looseness -1
``weak'' corrections
using $\theta_{\cut}=20^\circ$
for various values of the Higgs-boson mass. 
While the Higgs resonance is comparably sharp for small Higgs-boson masses,
it is washed out by the large width of the Higgs boson for high $\MH$.
Already for $\MH=400\GeV$ the Higgs resonance is hardly visible. 

\section{Summary}
 
The process $\AAWW$ will be one of the most interesting reactions at future
$\ga\ga$ colliders.
In particular, it is very useful to study non-Abelian gauge couplings.
 
We have calculated the one-loop radiative corrections to $\AAWW$ within the
electroweak Standard Model in the soft-photon approximation for arbitrary
polarizations of the photons and \PW~bosons. 
An interesting peculiarity of $\AAWW$
is the absence of most universal leading corrections, such as 
the running of $\al$ and leading logarithms associated with collinear
bremsstrahlung.
Therefore, theoretical predictions are very clean.
The variation of the \css\ with the top-quark and Higgs-boson masses
is small if $\MW$ is kept fixed with the exception of the \css\
involving longitudinal \PW~bosons at high 
energies. 
In the heavy mass limit no leading $\Mt^2$-,
$\log\Mt$-, and $\log\MH$-terms exist.

The soft-photon-cut-off-indepen\-dent radiative corrections to the
total \cs\ are of the order of 10\%
and can reach up to 50\% at $2\TeV$.
The large corrections are due to bosonic loop diagrams whereas the
effects of the fermionic ones are of the order of 5--10\%.

\end{document}